\documentclass[preprint,superscriptaddress,showpacs]{revtex4-1}
\usepackage{graphicx}
\usepackage{amsmath}

\begin{document}

\title{Shock-driven transition to turbulence: emergence of power-law scaling}

\author{D. Olmstead}
\affiliation{The University of New Mexico}
\author{P. Wayne}
\affiliation{The University of New Mexico}
\author{D. Simons}
\affiliation{The University of New Mexico}
\author{I. Trueba Monje}
\affiliation{The University of New Mexico}
\author{J.H. Yoo}
\affiliation{The University of New Mexico}
\author{S. Kumar}
\affiliation{Indian Institute of Technology -- Kanpur}
\author{C.R. Truman}
\affiliation{The University of New Mexico}
\author{P. Vorobieff}
\affiliation{The University of New Mexico}

\begin{abstract}
We consider two cases of interaction between a planar shock and a cylindrical density interface.
In the first case (planar normal shock), the axis of the gas cylinder is parallel to the shock
front, and baroclinic vorticity deposited by the shock is predominantly two-dimensional 
(directed along the axis of the cylinder). In the second case, the cylinder is tilted, resulting in 
an oblique shock interaction, and a fully three-dimensional shock-induced vorticity field. 
The statistical properties of the flow 
for both cases are analyzed based on images from two orthogonal visualization planes, using structure functions of the intensity maps of fluorescent tracer pre-mixed with the heavy gas. 
At later times, these structure functions exhibit power-law-like behavior over a considerable range 
of scales. Manifestation of this behavior is remarkably consistent in terms of dimensionless 
time $\tau$ defined based on Richtmyer's linear theory within the range of Mach numbers 
from 1.1 to 2.0 and the range of gas cylinder tilt angles with respect to the plane of the shock 
front (0 to 30$^\circ$).
\end{abstract}


\maketitle

Richtmyer-Meshkov instability (RMI) \cite{richtmyer,meshkov} develops on an impulsively accelerated density 
interface, often manifesting in shock-accelerated gases with density gradients. RMI is 
responsible for vortex formation in a number of problems, from astrophysical \cite{rmi-supernova} and geophysical 
\cite{rmi-geophys} to such engineering applications as inertial confinement fusion \cite{rmi-icf} 
and supersonic combustion \cite{yang}. 

The growth of RMI is usually described in a sequence of stages \cite{brouillette, kumar}.  During the initial (linear) stage, 
the growth is to some extent consistent with Richtmyer's original theory \cite{richtmyer}, and is well described with compressible
linear theory \cite{yang94}. As initial vorticity deposition due to the misalignment of pressure and density gradients leads to 
roll-up of vortices, the second stage of nonlinear, vorticity-dominated deterministic growth follows \cite{zhangsohn,sadot}. At the same time, secondary instabilities due to shear and secondary baroclinic effects emerge, leading to
the next stage, where deterministic and disordered flow features coexist, and eventually to turbulence. 

Consider an interfacial perturbation with a
wavelength $\lambda$ (with the corresponding wavenumber $\kappa=2\pi / \lambda$) and amplitude $a_0$.
Let the interface initially 
separate gases of densities $\rho_1$ and $\rho_2$ and be accelerated with a shock of Mach number $M$.
The perturbation growth will depend in a non-trivial way on $M$, the Atwood number 
$A=(\rho_2 - \rho_1)/ (\rho_2+ \rho_1)$, on initial interfacial geometry, on the extent of diffusion at the interface, etc. In the simplest case (Richtmyer's linear theory for a sharp periodically perturbed interface), the perturbation growth rate can be described as 
\begin{equation}
\label{richt}
|v^{imp}| = \kappa a_0 A \Delta V
\end{equation}
 where $\Delta V$ is the piston velocity of the shocked flow, dependent on the Mach number. 
Unlike the related Rayleigh-Taylor instability, which is gravity-driven and thus has a constant supply of energy, the energy provided to produce the growth of RMI is finite, 
and thus the growth generally slows with time. Several models exist to describe the RMI growth rate, from well-considered theories \cite{sadot} to semi-empirical equations \cite{rightley99}. 

During the subsequent stage of evolution, the RMI-driven mixing flow is known to 
develop features consistent with transition to turbulence: greatly enhanced mixing 
(mixing transition \cite{rightley99}), power-law scalings of structure functions
of scalar tracer advected by the flow \cite{vorobieff98} and of velocity \cite{mohamed} consistent with classical predictions for fully-developed turbulence \cite{obukhov,corrsin,k41}. 

Until recently, experimental studies of RMI-driven transition to turbulence were largely confined
to a situation when the initial conditions are nominally two-dimensional (2D), leading to formation of large-scale vortices with vorticity initially confined to one direction. Here we present a comparative study of shock-driven transition to turbulence evolving from such nominally 
two-dimensional conditions and from conditions when initial vorticity deposition is inherently 
three-dimensional, and find that the scalings of emerging turbulence are remarkably consistent for 
the geometries we investigate.  

\begin{figure}
\centerline{\includegraphics{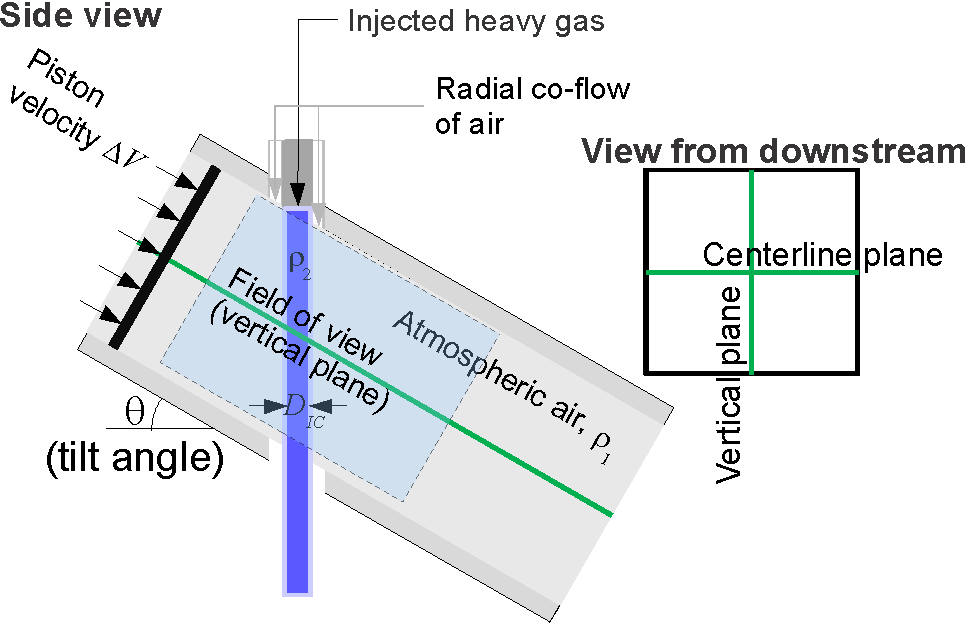}}
\caption{\label{setup} Experimental setup. Top -- side view, bottom -- view from downstream showing centerline and vertical visualization planes.}
\end{figure}

The experiments described here were conducted at the UNM shock tube \cite{prl,anderson12} 
(Fig.~\ref{setup}). Gravity-driven flow through a cylindrical nozzle produced the initial conditions, with a diffuse interface forming between a heavy gas (mixture of 89\% SF$_6$ and 11\% of acetone tracer by volume) injected through the nozzle into a test section of the shock tube, 
the latter being filled with quiescent air at ambient pressure. The measured  
Atwood number characterizing this diffuse interface is 0.6.
Planar laser-induced fluorescence (PLIF) is induced in acetone tracer by illuminating a 
planar section of the flow with a pulsed UV laser sheet at a wavelength 266~nm. 
A distinctive feature of the UNM shock tube is that it can be tilted with
respect to the horizontal by an angle $\theta$, making it possible to create initial conditions
both for planar normal (quasi-2D) and oblique shock interaction with density interfaces. 

In the experiments described here, the density interface was produced by injecting SF$_6$ with
acetone tracer through a cylindrical nozzle (diameter $D_{IC}=6.35$~mm). The heavy gas flow was stabilized by concentric co-flow of air, resulting in a highly repeatable laminar diffuse interface. The shock tube was tilted at angles $\theta = 0$, 20$^\circ$ and 30$^\circ$ with
the horizontal. Evolution of the flow was visualized in two planes -- the vertical plane and the
plane of symmetry of the shock tube cross-section normal to the vertical. At $\theta = 0$ the
second plane would be horizontal, at other angles, it is tilted with the shock tube. We refer to this second plane of visualization as the centerline plane. Images in the visualization plane were captured with 
a 4-megapixel backward-oriented and cooled CCD camera with 16 grayscale bits per pixel and a quantum
efficiency about 90\%.

\begin{figure}
\centerline{\includegraphics[width=6.325in]{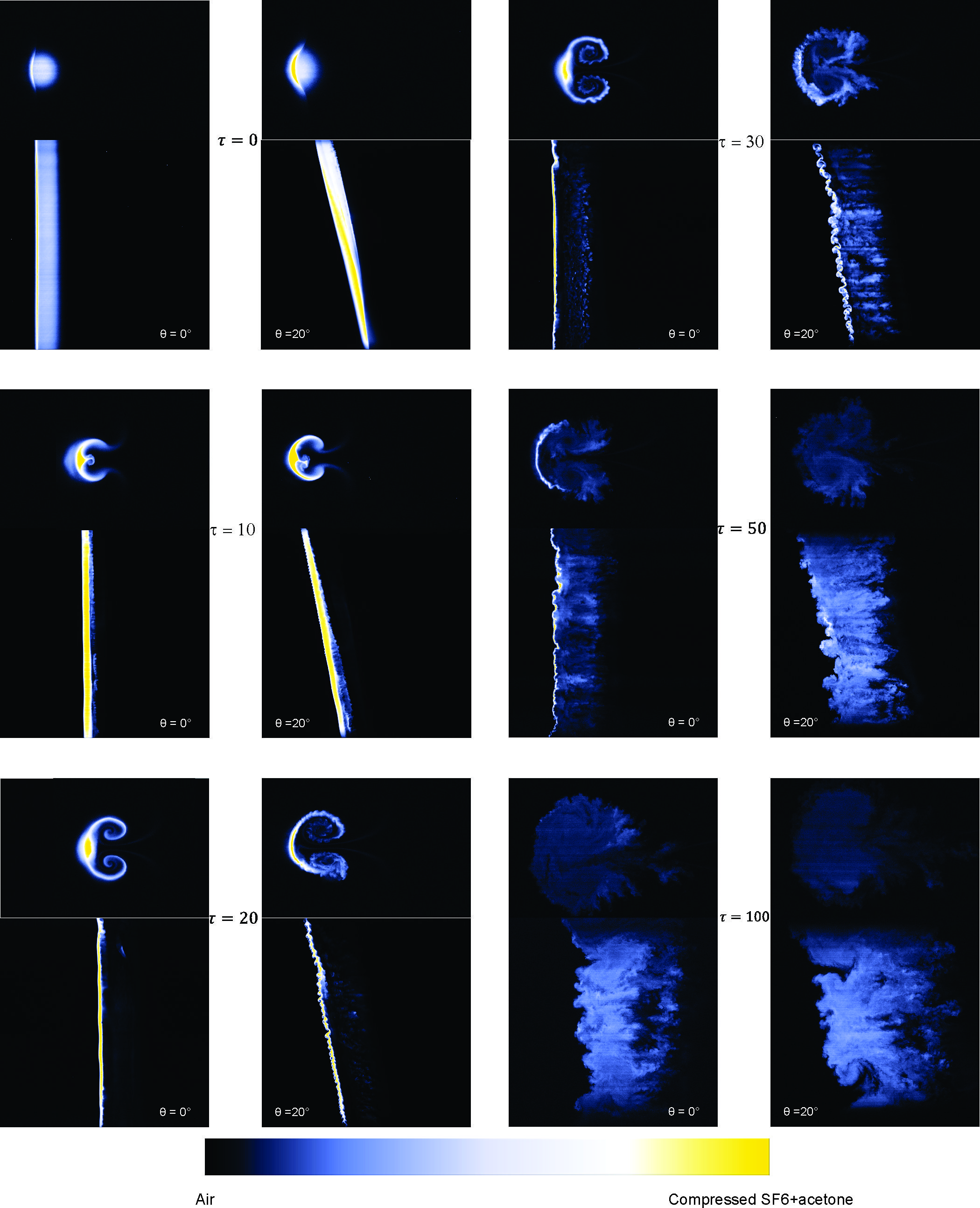}}
\caption{\label{pretty} Shock-accelerated gas cylinder evolution at $M=2$, $A=0.6$ 
for tilt angles $\theta = 0$ and $20^\circ$. Images in the centerline (above) and vertical (below)
planes are paired. Dimensionless time $\tau$ is labeled. 
Color is artificial, streamwise image extent is 44~mm.}
\end{figure}

Figure \ref{pretty} shows a comparison of flows evolving from a quasi-2D initial conditions 
at $\theta=0$ and from three-dimensional initial conditions at $\theta=20^\circ$. The dimensionless time used to label the images is $\tau=k A\Delta V(t-t_0)$, where $k$ and 
$\Delta V$ are the wavenumber and the piston velocity introduced earlier. For a cylinder of diameter $D$, $k= 2\pi / D$. Time $t = t_0$ corresponds to the shock traversing the center of the initial column. 
In terms of timing $\tau$, the initial linear growth rates (Eq.~\ref{richt}) would remain the same 
for the same geometry of the initial conditions, not changing with  $A$  or $M$ 
(or, to be more exact, $\Delta V$).  

In both cases ($\theta = 0$ and $\theta = 20^\circ$), for a substantial time, 
the flow in the centerline plane is 
dominated by a pair of counter-rotating vortex columns well-known from earlier studies. 
However, flow evolution in the vertical plane is different -- for oblique shock interaction,
vorticity of the same sign is deposited along the oblique density interfaces, leading to 
formation of shear layer-like structures, emerging at $\tau=10$ and clearly 
apparent at $\tau=20$. 

How can we quantify the apparent transition to turbulence here? 
Statistically, turbulence has long been associated with power-law behavior of spectra
within the inertial range. 
It is important to note that, while the spectral representation of Kolmogorov theory 
(``the five thirds law'' \cite{kolmogorov1962refinement}) is perhaps better known, 
the original 1941 paper (K41 \cite{k41}) 
 actually dealt with real-space properties of turbulence based on two-point velocity
correlations and the statistics of velocity structure functions.

\begin{figure*}[!htbp]
\centerline{\includegraphics{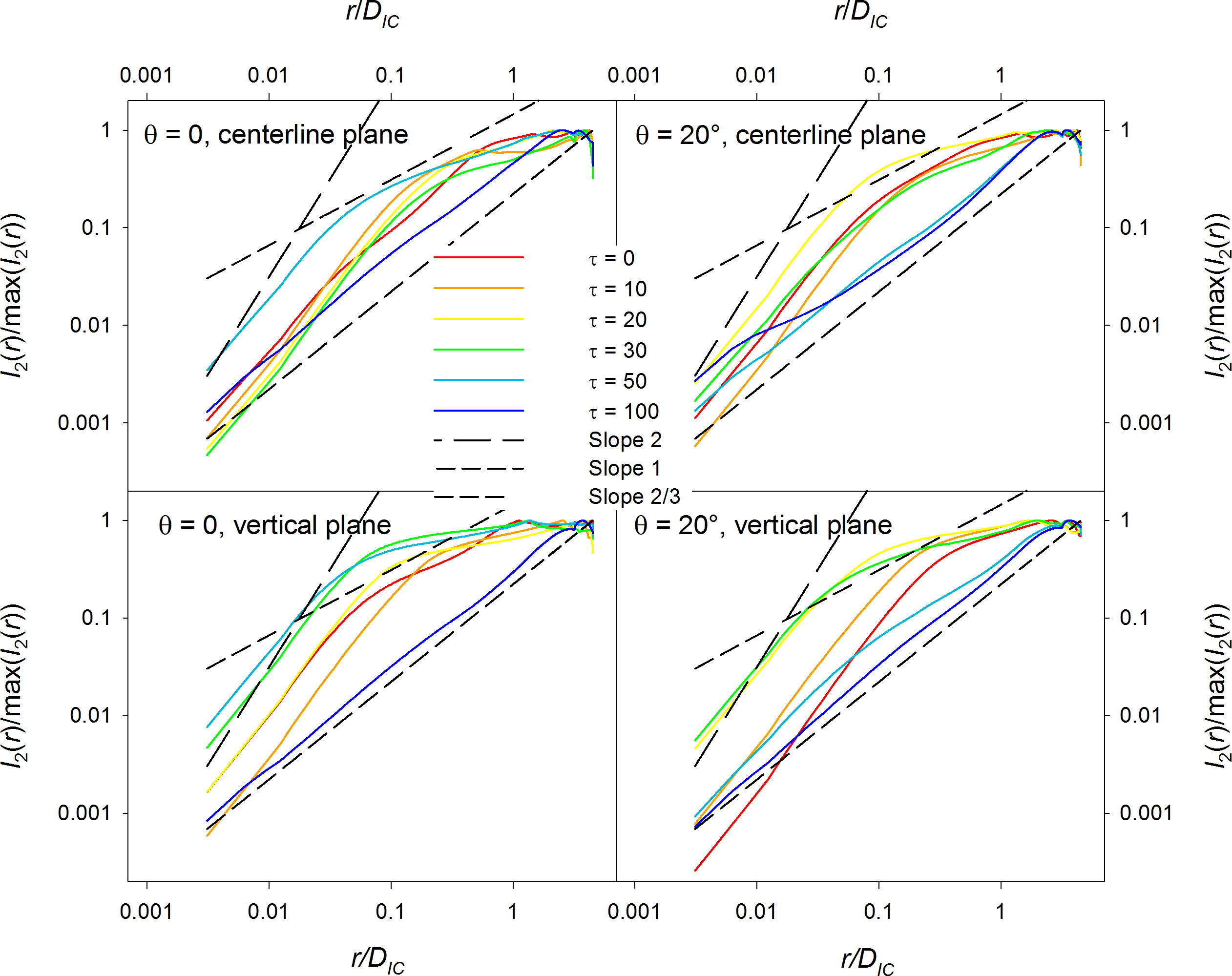}}
\caption{\label{fig2} Second-order structure functions of fluorescence intensity $I_2(r)$ in images shown
in Fig.~\protect\ref{pretty}. The values of the structure functions are normalized by their maxima, $r$ 
is scaled by the injection nozzle diameter $D_{IC}$. The plots are color-coded by dimensionless times $\tau$.
Power-law scalings with exponents 2, 1, and 2/3 are shown as guides to the eye.
}
\end{figure*}

Velocity structure function evolution in a shock-accelerated heavy-gas cylinder flow 
was studied for small ($M=1.2$) Mach numbers \cite{mohamed} with 
particle image velocimetry (PIV). The late-time results were roughly consistent with K41
prediction of 2/3 power-law scaling for the second-order longitudinal velocity 
structure function. 
Sadly, at
higher Mach numbers, tracer particles used for PIV present an increasing problem because 
they
don't follow the gas flow \cite{anderson2015experimental} and interfere with flow 
physics \cite{prl}. Here we use a cleaner 
diagnostic (PLIF), however, it does not easily yield results in terms of velocity, because it effectively 
shows cross-sections of a scalar field (fluorescence intensity is related to tracer concentration and
thus to local volume fraction of injected gas mixture) 
advected by the flow. In turbulent flow, such scalar fields are long known to develop 
power-law statistics as well. Corrsin \cite{corrsin} famously derived the equation for the
spectrum of temperature fluctuations in isotropic fully developed turbulent flow with $k^{-5/3}$ exponent. This result can be generalized to the spectrum of any diffusive passive scalar and even of a reacting component in the flow \cite{monin2013statistical}, and moreover, has an equivalent 
representation in terms of the second-order structure function of the scalar (under the same
conditions that ensure the equivalence of the -5/3 and 2/3 laws for velocity spectra and structure
functions \cite{monin2013statistical}). The second-order structure function of the scalar 
in fully developed turbulent flow should thus also scale as $r^{2/3}$. Power-law scaling emergence 
for a Mie scattering intensity field from submicron-sized particles pre-mixed with a varicose 
curtain of heavy gas 
was reported \cite{vorobieff98} for a low Mach number (1.2) shock-driven flow over a range of more than 
one order of magnitude, with a 0.73 exponent. 

Advances in image acquisition and experimental techniques now make it possible to resolve the entire
range of physically relevant scales in laboratory shock-driven flow -- from energy injection scale 
(centimeter-sized baroclinically produced vortices) down to Kolmogorov 
length scale (order of microns). With 
fluorescent gas as tracer, flow tracking fidelity also ceases to be a problem at higher Mach numbers.
Figure~\ref{fig2} shows plots of the second-order structure function $I_2(r)$ of fluorescence intensity 
$I$ of the 
tracer  after $M=2$ shock acceleration, which should map the local concentration of the injected gas cylinder 
material (SF$_6$ with acetone):
$
I_2(r)=\langle \left(I(\mathbf{x})- I(\mathbf{x} + \mathbf{r}) \right)^2 \rangle
$.
Here $\langle \cdot \rangle$ denotes ensemble-averaging over all pairs of points in the image separated 
by a distance $r = \left|\mathbf{r}\right| $.
 On the time scale of the experiments, differential diffusion of acetone
and SF$_6$ plays no role.

\begin{figure}
\centerline{\includegraphics{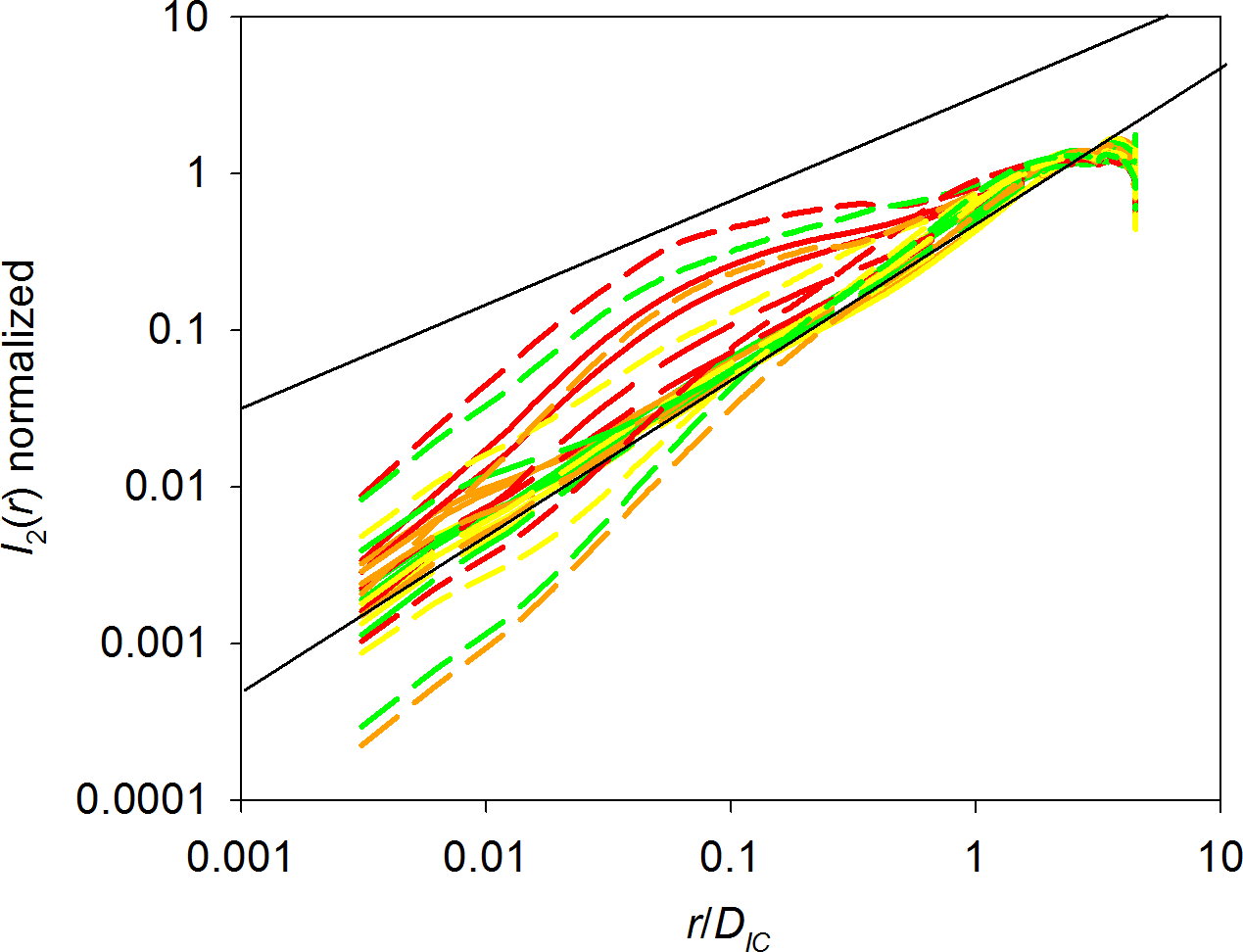}}
\caption{\label{late-times} Late-time ($\tau \gtrsim 100$) 
structure functions of fluorescence intensity $I_2(r)$ normalized by their average values for 
each image. Mach numbers are color-coded: red - 1.1, 
orange - 1.4, yellow - 1.7, green - 2.0. Solid lines denote $\theta=0$, long dash -- 20$^\circ$, 
short dash -- 30$^\circ$. Thin black lines shows slopes 2/3 (top) and 1 (bottom).}
\end{figure}

Significant differences are apparent in the flow morphology at $\theta=0$ and $\theta=20^\circ$, 
especially in the vertical plane. These differences and the presence 
of notable anisotropy notwithstanding,
the late-time ($\tau \sim 100$) behaviors of $I_2(r)$ are remarkably similar and close to a power law. What is 
also noteworthy is that the exponent of best power-law fit to late-time $I_2(r)$ is appreciably higher than
the fully developed turbulence prediction (2/3) -- in fact, it's close to unity. Moreover, this late-time 
behavior is quite prominent in the parameter space we investigated: $M = 1.13$, 1.4, 1.7, 2.0, 
$\theta=0$, 20$^\circ$, 30$^\circ$, manifesting over at least a decade in 17 out of 24 plots in 
Fig.~\ref{late-times}. 

In the context of transition to turbulence, formation of a cascade with power-law scaling of the structure functions is expected. In a sense, similar behavior across a range of different initial geometries and 
Mach numbers is also consistent with the notion of transition to turbulence, when the flow ``forgets'' its
initial conditions. A question that arises, however, is why $I_2(r)$ scales as $r^1$ rather than as $r^{2/3}$. 
Some of the likely answers are that the scaling emerges in flow that does not fit the definition of 
fully developed turbulence -- it is transitional, driven by a finite energy input, and significantly 
anisotropic even at late times. Scalar spectra are well-known to deviate from the Obukhov-Corrsin value
\cite{celani}, and this applies even to flows where the velocity field scaling is consistent with 
K41 theory predictions \cite{sreenivasan96,innocenti2001}. The latter reference is of specific 
interest because it deals with the transformation of a ``blob'' of scalar initially injected in the
flow, which is not very far from our visualization scheme.

Velocity-field isotropy is very important
for the Obukhov-Corrsin scalar scaling to manifest \cite{sreenivasan96}, because in most estimates 
of scalar dissipation, the assumption of local isotropy is used. Accordingly, significant deviation of
scalar scaling from the 2/3 value is notable for shear flows \cite{sreenivasan96,celani}. 
In the flow under consideration here, shear plays a major role -- both in formation of secondary instabilities
in the centerline plane and in the apparent Kelvin-Helmholtz vortex formation in the vertical plane. 
To the best of our knowledge, the scalar structure function scaling we observe has not been reported 
previously, and, while not totally physically unexpected, is quite interesting and deserving further study. 
 
\noindent This work is supported by the US Department of Energy grant DE-NA-0002913.

\bibliography{prl15}
\end{document}